\documentclass[aps,prb,twocolumn,amsmath,nofootinbib]{revtex4-1}

\usepackage{epsfig}
\usepackage[colorlinks,linkcolor=blue,anchorcolor=blue,
            citecolor=blue,urlcolor=blue,
            breaklinks=true]{hyperref}
\usepackage{graphics}
\usepackage{color}
\usepackage{dcolumn}
\usepackage{bm}
\usepackage{cases}
\usepackage{appendix}
\usepackage{lineno,hyperref}
\usepackage{ulem}

\newcommand{\tb}{\textbf}

\begin{document}
\author{Wei-Ran Cao$^{1}$}
\author{Yong-Long Wang$^{1,2}$}
 \email{wangonglong@lyu.edu.cn}
\author{Xiao-Lei Chen$^{3}$}
\author{Hua Jiang$^{1}$}
\author{Chang-Tan Xu$^{1}$}
\author{Hong-Shi Zong$^{2,4,5}$}
\email{zonghs@nju.edu.cn}
\address{$^{1}$ School of Physics and Electronic Engineering, Linyi University, Linyi 276005, P. R. China}
\address{$^{2}$ Department of Physics, Nanjing University, Nanjing 210093, P. R. China}
\address{$^{3}$ Linyi Radio Frequency and Station Management, Linyi 276000, P. R. China}
\address{$^{4}$ Joint Center for Particle, Nuclear Physics and Cosmology, Nanjing 210093, P. R. China}
\address{$^{5}$ State Key Laboratory of Theoretical Physics, Institute of Theoretical Physics, CAS, Beijing 100190, P. R. China}

\title{The geometric potential of a double-frequency corrugated surface}
\begin{abstract}
For an electron confined to a surface reconstructed by double-frequency corrugations, we give the effective Hamiltonian by the formula of geometric influences, obtain an additive scalar potential induced by curvature that consists of attractive wells with different depth. The difference is generated by the multiple frequency  of the double-frequency corrugation. Subsequently, we investigate the effects of geometric potential on the transmission probability, and find the resonant tunneling peaks becoming rapidly sharper and the transmission gaps being substantially widened with increasing the multiple frequency. As a potential application, double-frequency corrugations can be employed to select electrons with particular incident energy, as an electronic switch, which are more effective than a single-frequency ones.
\bigskip

\noindent PACS Numbers: 73.50.-h, 73.20.-r, 03.65.-w, 02.40.-k
\end{abstract}
\maketitle

\section{INTRODUCTION}
With the rapid development of nanotechnology, the fabrication of electronic nanodevices with complex geometries is becoming easy, and therefore various two-dimensional (2D) nanosystems with corrugations are designed and implemented, such as corrugated films~\cite{Ono2009Tuning, Ortix2011Curvature, Kosugi2011Pauli, Zhang2015A, Wang2016Transmission}, bent and corrugated nanotubes~\cite{Gupta2011Geometrical, Goldstone1992Bound, Shima2010Diverse, Novakovic2011Transport, Cheng2018Geometric} and so on. Those successes in manufacture found the basis of investigating the curvature-tunable filter~\cite{Wang2016Transmission}. Those surfaces with corrugations are an important type of 2D systems with relative complex geometries. For an electron confined to a corrugated surface, the effective Hamiltonian does not only depend on the 2D intrinsic space geometry, but also on the embedding of the surface in three-dimensional (3D) Euclidean space~\cite{Jaffe2003Quantum}. To deduce the effective Schr\"{o}dinger equation, the thin-layer quantization formalism~\cite{Jensen1971Quantum, Costa1981Quantum} is a suitable approach. Experimentally, the geometric potential was realized in photonic topological crystals~\cite{Szameit2010Geometric}, and the geometric momentum~\cite{Liu2007Constraint, Liu2011Geometric, Liu2013Geometric, Wang2017Geometric} was observed to govern the propagation of surface plasmon polaritons on metallic wires~\cite{Schmidt2015Curvature}. The two evidences demonstrate the validity of the thin-layer quantization scheme.

With the development of the thin-layer quantization theory~\cite{Jensen1971Quantum, Costa1981Quantum, Wang2014Pauli, Wang2016Quantum}, quantum electronic devices are rapidly developing toward small size and complex structure and geometry. When the size of electronic devices is small enough to be compared with de Broglie wavelength, the geometric effects will become very significant~\cite{Wang2016Transmission}. For investigating the effects of geometric potential, the nanoscale corrugation was introduced into a thin film to study the related electronic properties~\cite{Ono2009Tuning}, they obtained the stepwise resistivity determined by the corrugation. In curved quantum waveguides, the bound states were obtained~\cite{Exner1989Bound, Duclos1995Curvature, Bulla1997Weakly, Griffiths2001Waves}, and the related transmission were investigated~\cite{Porod1992Transmission, Porod1993Resonance, Porod1994Transmission}, too. Recently, we found that transmission gaps and resonant tunneling domains produced by corrugation~\cite{Wang2016Transmission}. To the best of our knowledge, most of the investigations have been done on single-frequency corrugation, and rarely involve double-frequency or multiple-frequency corrugations. However, the double-frequency potentials were recently employed to describe one-dimensional superlattice~\cite{Citro2016A}, and 2D superlattice potential~\cite{Lohse2018Exploring}. As a consequence, the geometric effects of double-frequency corrugation need a further investigation.

In the present paper, we will consider a model including double-frequency corrugations, and investigate the influences of the multiple frequency on the geometric potential, and on the transmission probability. The double-frequency corrugation can contribute a list of attractive wells with different depth. When the multiple frequency increases, the resonant tunneling peaks rapidly become sharper, and the transmission gaps effectively become wider. As a potential application, the results are useful for designing quantum-electromechanical circuits~\cite{Blencowe2004Quantum, Chaplik2004On, Xiang2013Hybrid} and thin film transistors~\cite{Amalraj2014Influence}.

This paper is organized as follows. In Sec.~\ref{2}, using the formula of geometric influences~\cite{Wang2017Geometric, Wang2018Geometric, Wang2018Erratum}, we obtain the effective Hamiltonian for an electron confined to a thin film reconstructed by double-frequency corrugation and the geometric potential. In Sec.~\ref{3}, we investigate the effects of the double-frequency corrugation on transmission probability, especially the multiple frequency. Finally in Sec.~\ref{4} the conclusions are given.

\section{Quantum dynamics of a particle confined on a periodically corrugated surface}\label{2}
In quantum mechanics, a free microscopic particle can be described by a time-independent Schr\"{o}dinger equation, that is
\begin{equation}\label{OSE}
-\frac{\hbar^2}{2m^*}\nabla^2\psi=E\psi,
\end{equation}
where $\hbar$ is the Plank constant divided by $2\pi$, $m^*$ is the effective mass of particle, $\nabla^2$ is the Laplace operator usually defined in a 3D flat spatial space, $\psi$ is the wave function describing the motion of particle and $E$ is the energy with respect to $\psi$.

When the free particle is confined to a curved surface, the effective dynamics would be affected by the geometrical properties~\cite{Costa1981Quantum, Wang2017Geometric}. In the spirit of the thin-layer quantization scheme, to obtain the effective quantum dynamics describing a particle confined to a curved surface, the Schr\"{o}dinger equation Eq.~\eqref{OSE} should be originally defined in a 3D curvilinear coordinate system as below
\begin{equation}\label{CSE}
-\frac{\hbar^2}{2m^*}\frac{1}{\sqrt{G}}\partial_iG^{ij}\sqrt{G}\partial_j\psi=E\psi,
\end{equation}
where $G$ and $G^{ij}$ are the determinant and the inverse of the metric tensor $G_{ij}$ defined in a 3D curvilinear coordinate system, respectively, wherein $i, j=1,2,3$ denote the three curvilinear coordinate variables. It is obvious that the metric $G_{ij}$ contained in the Laplace operator depends on $q_3$, the coordinate variable normal to a curved surface. The $q_3$ dependence contributes a scalar potential in the effective Hamiltonian.

\begin{figure}[htbp]
  \centering
  \includegraphics[width=0.44\textwidth]{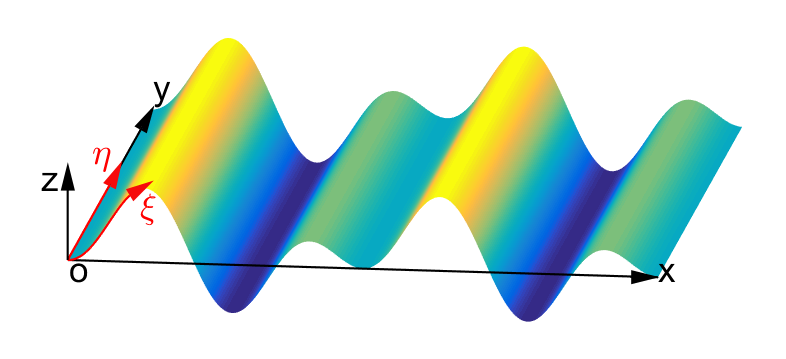}
  \caption{\footnotesize (Color online) Schematic of a surface reconstructed by double-frequency corrugations, $z=a\cos(n\gamma x)\sin(\gamma x)$. Here $a$ and $2\pi/\gamma$ are the amplitude and period length of corrugation, respectively. $n$ is a nonzero positive integer. $(\xi,\eta)$ denotes the two curvilinear coordinate variables over surface.}\label{corrugation}
\end{figure}

A surface reconstructed by double-frequency corrugations $\mathcal{S}$ (see Fig.\ref{corrugation}) that can be parametrized by
\begin{equation}\label{csurf}
z=a\cos(n\gamma x)\sin(\gamma x),
\end{equation}
where $a$ denotes the amplitude of composite corrugation, $2\pi/\gamma$ describes the period length of corrugation, $n$ is a nonzero positive integer which is named as a multiple frequency because that it describes the high frequency composition of the double-frequency corrugation. In the composite corrugation Eq.~\eqref{csurf}, $\cos(n\gamma x)$ describes the high frequency composition and $\sin(\gamma x)$ does the low frequency one. The position vector of a point on $\mathcal{S}$ can be then described by $\tb{r}=x\tb{e}_x+y\tb{e}_y+a\cos(n\gamma x)\sin(\gamma x)\tb{e}_z$, and hence that of a point in the immediate region near to $\mathcal{S}$ can be parameterized by $\tb{R}=\tb{r}+q_3\tb{e}_n$, $\tb{e}_n$ is the unit vector basis normal to $\mathcal{S}$, and $q_3$ is the corresponding coordinate variable. According to the definitions $G_{ij}=\partial_i\tb{R}\cdot\partial_j\tb{R}$ and $g_{ab}=\partial_a\tb{r}\cdot\partial_b\tb{r}$, wherein $(i,j=1,2,3)$ and $(a,b=1,2)$, we obtain
\begin{equation}\label{SurfMetric}
g_{ab}=
\left (
\begin{array}{cc}
1 & 0\\
0 & 1
\end{array}
\right )
\end{equation}
and
\begin{equation}\label{VMetric}
G_{ij}=
\left (
\begin{array}{ccc}
f^2 & 0 & 0\\
0 & 1 & 0\\
0 & 0 & 1
\end{array}
\right ),
\end{equation}
respectively. It is easy to obtain the relationship between $G$ and $g$, $G=f^2g$, where the rescaling factor $f$ is
\begin{widetext}
\begin{equation}\label{factor}
f=1+\frac{a[(1+n^2)\cos n\gamma x\sin\gamma x+2n\cos\gamma x\sin n\gamma x]}{[1+a^2(\cos\gamma x\cos n\gamma x-n\sin\gamma x\sin n\gamma x)^2]^{3/2}}q_3.
\end{equation}

According to the above results, using the formula of the geometric influence~\cite{Wang2017Geometric,Wang2018Geometric, Wang2018Erratum}, we can calculate the effective Hamiltonian as
\begin{equation}\label{EffH}
\rm{H}_{eff}=\langle\chi_{0_n}|f^{\frac{1}{2}}\rm{\hat{H}} f^{-\frac{1}{2}}-\rm{\hat{H}}_n|\chi_{0_n}\rangle_0
=-\frac{\hbar^2}{2m^*}(\partial_{\eta}^2+\partial_{\xi}^2)+V_g,
\end{equation}
where $\rm{\hat{H}}$ is the original Hamiltonian describing a free particle in a 3D curvilinear coordinate system with $\rm{\hat{H}}=-\frac{\hbar^2}{2m^*}\frac{1}{\sqrt{G}}\partial_i\sqrt{G}G^{ij}\partial_j$, $\rm{\hat{H}}_n$ denotes the normal component of $\rm{\hat{H}}$ with $\rm{\hat{H}}_n=-\frac{\hbar^2}{2m^*}\frac{\partial^2}{\partial q_3^2}$, and $V_g$ is the well-known geometric potential~\cite{Jensen1971Quantum, Costa1981Quantum} that reads
\begin{equation}\label{GP}
V_g=-\frac{\hbar^2}{2m^*}\frac{a^2\gamma^4[(1+n^2)\cos n\gamma x\sin\gamma x+2n\cos\gamma x\sin n\gamma x]^2}{4[1+a^2\gamma^2(\cos\gamma x\cos n\gamma x-n\sin\gamma x\sin n\gamma x)^2]^3}.
\end{equation}
\end{widetext}
In the previous calculations, the wave function of the ground state $|\chi_{0_c}\rangle$ is taken as the ground state of a harmonic oscillator with $w\to\infty$, that is
\begin{equation}
|\chi_{0_n}\rangle=\alpha^{\frac{1}{2}}\pi^{-\frac{1}{4}}e^{-\frac{\alpha^2q_3^2}{2}},
\end{equation}
where $\alpha=\sqrt{\frac{mw}{\hbar}}$. The evidence is that the geometric potential does not depend on the specific form of the confining potential and its relative strength~\cite{Ortix2015Quantum, Encinosa2005Wave}. By virtue of the effective Hamiltonian Eq.~\eqref{EffH}, we directly write the effective Schr\"{o}dinger equation as
\begin{equation}\label{ESE}
-\frac{\hbar^2}{2m^*}(\frac{\partial^2}{\partial\xi^2}+\frac{\partial^2}{\partial\eta^2}) \chi_s+V_g\chi_s=E_s\chi_s,
\end{equation}
where $\eta$ and $\xi$ are two tangent coordinate variables of $\mathcal{S}$, respectively. Due to $V_g$ only depending on $x$, and $\xi$ just being a function of $x$, we can rewrite $\chi_s=\chi_{\eta}(\eta)\chi_{\xi}(\xi)$, and separate the effective quantum equation analytically into two components of $\eta$ and $\xi$, they are
\begin{equation}\label{XiSE}
-\frac{\hbar^2}{2m^*}\frac{\partial^2}{\partial\xi^2}\chi_{\xi}+V_g\chi_{\xi} =E_{\xi}\chi_{\xi}
\end{equation}
and
\begin{equation}\nonumber
-\frac{\hbar^2}{2m^*}\frac{\partial^2}{\partial\eta^2}\chi_{\eta}=E_{\eta}\chi_{\eta},
\end{equation}
respectively, where $E_s=E_{\xi}+E_{\eta}$.

\begin{figure}[htbp]
  \centering
  \includegraphics[width=0.49\textwidth]{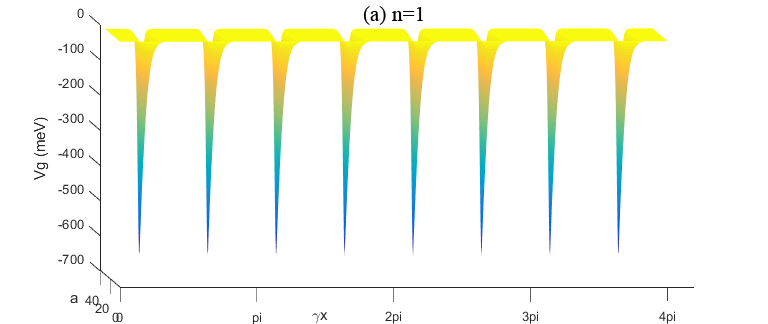}\\
  \includegraphics[width=0.49\textwidth]{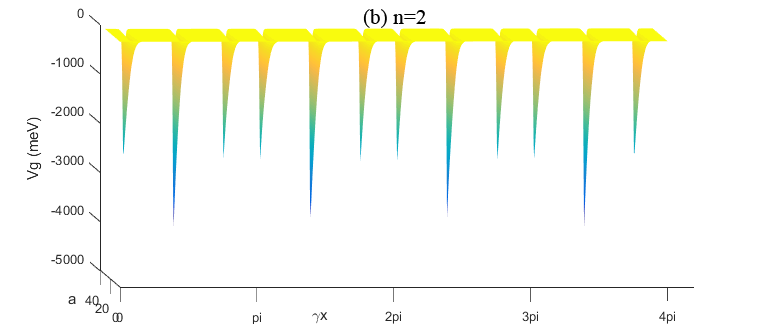}\\
  \includegraphics[width=0.49\textwidth]{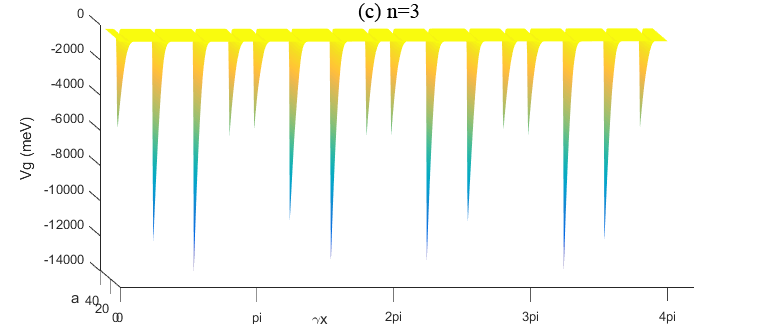}\\
  \includegraphics[width=0.49\textwidth]{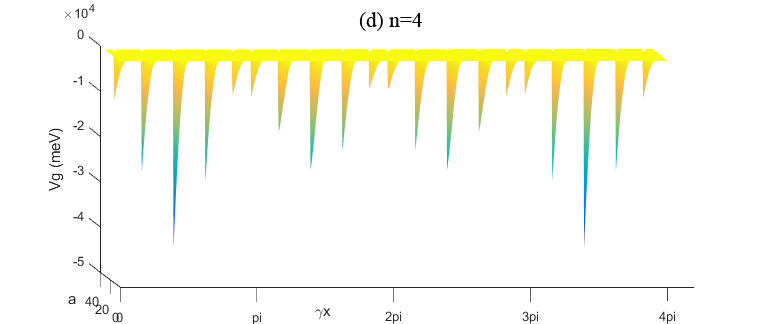}
  \caption{\footnotesize (Color online) Surfaces of $V_g$ as function of $\gamma x$ and $a$, $a$ ranging from $0$ to $30nm$ and $\gamma x$ varying in the region $[0, 8\pi]$, for (a) $n=1$, (b) $n=2$, (c) $n=3$ and (d) $n=4$.}\label{GPFig2}
\end{figure}

It is worthwhile to notice that in the original Schr\"{o}dinger equation Eq.~\eqref{CSE} the metric tensor $G_{ij}$ depends on $q_3$, while in the effective Schr\"{o}dinger equation the metric $g_{ab}$ does not. In the thin-layer quantization procedure, the $q_3$ dependence of the metric $G_{ij}$ contributes a geometric potential to keep the consistency of the effective Schr\"{o}dinger equation. As a central result of contributions of the thin-layer quantization scheme, the geometric potential is presented in the effective Hamiltonian and plays an important role in the effective dynamics. The geometric potential is induced by curvature, therefore it can be reconstructed by designing the geometry of the investigated system. In the present model, the most difference in corrugation is that it is determined by two frequencies compared to the usual single-frequency one.

To learn the actions of composite corrugation on the geometric potential, the geometric potential is shown in Fig.~\ref{GPFig2} as a function of $\gamma x$ and $a$ with $m^*=0.067m_0$, $m_0$ is the static mass of electron. In this case, the substrate of the composite corrugation is GaAs. Obviously, in the form the result shown in Fig.(a) is in nice agreement with that given by our another paper~\cite{Wang2016Transmission}. However, the depthes of potential wells produced by the double-frequency corrugation are particularly much more than that of wells given by the single-frequency one. When the multiple frequency number $n$ has a large integer, such as $n=2,3,4$, there is an obvious feature that is the attractive wells with different depthes (see Fig.~\ref{GPFig2} (b), (c) and (d)). It is worthwhile to notice that the depth differences are closely related to the multiple frequency. It is shown that the corrugation can substantially affect the geometric potential. As a consequence, we can construct particular wells array by introducing a composite corrugation, tuning its amplitude $a$, and adjusting the multiple frequency number $n$.

\section{Transmission probability affected by a composite corrugation}\label{3}
\begin{figure}[htbp]
\centering
\includegraphics[width=0.51\textwidth]{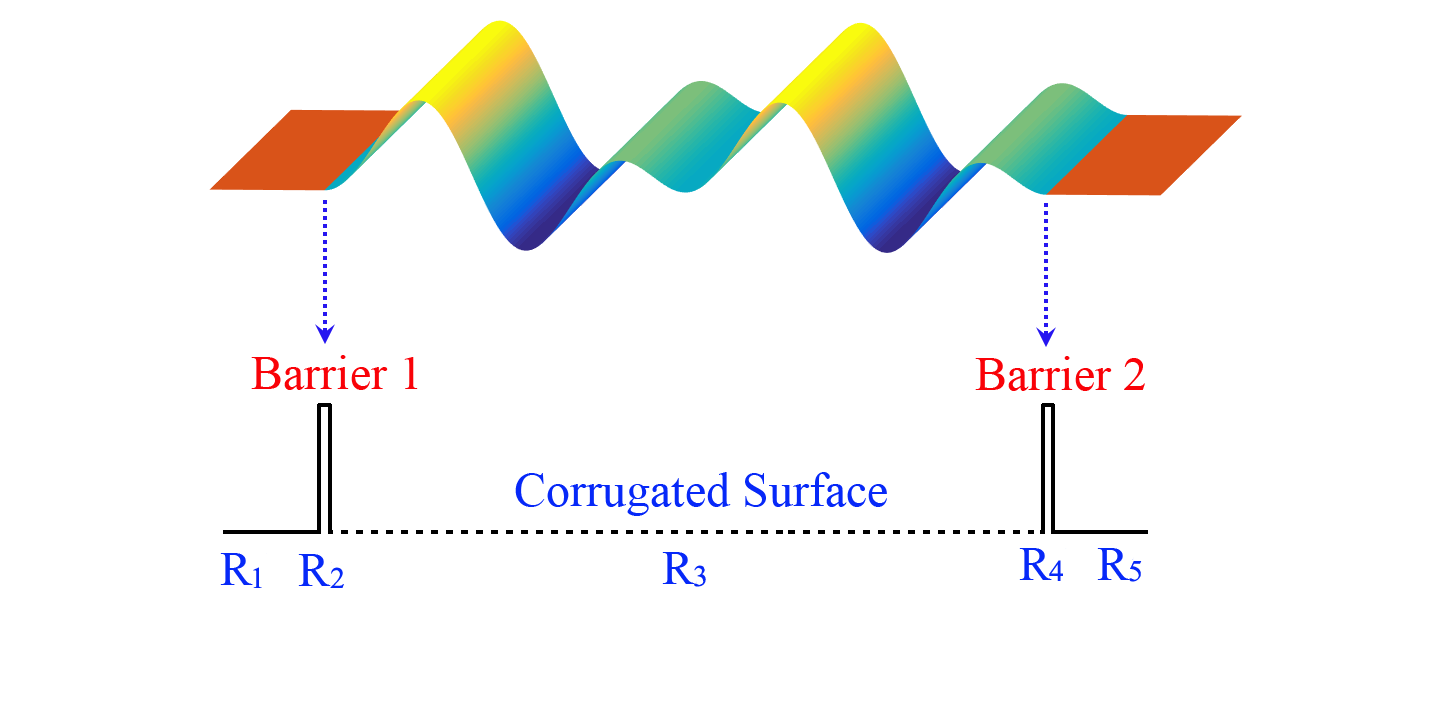}
\caption{\footnotesize (Color online) Schematic of the simplified model. $R_1$ and $R_5$ correspond to two leads, $R_2$ and $R_4$ describes the two potential barriers resulting from the boundaries between adjacent regions in which an electron has different effective masses, $R_3$ is a surface reconstructed by a composite corrugation.}\label{Model}
\end{figure}
From Eq.~\eqref{XiSE} and the surface $\mathcal{S}$ Eq.~\eqref{csurf}, the effective Schr\"{o}dinger equation can be reexpressed in $x$ as
\begin{equation}\label{4SE1}
-\frac{\hbar^2}{2m^*}\frac{1}{\Delta}\frac{d}{dx}[\frac{1}{\Delta} \frac{d}{dx}\psi(x)] +V(x)\psi(x)=E\psi(x),
\end{equation}
where $m^*=0.067m_0$, $m_0$ is the static mass of an electron, if $x\in R_3$ and $m^*=m_0$ otherwise,
\begin{equation}\nonumber
\Delta=\sqrt{1+a^2\gamma^2(\cos n\gamma x\cos\gamma x-n\sin n\gamma x\sin\gamma x)^2}
\end{equation}
if $x\in R_3$ and $\Delta=1$ otherwise, $\psi$ is a wave function, $E$ is the energy with respect to $\psi$, and $V(x)$ stands for the potential of the simplified model (see Fig.\ref{Model}), that is
\begin{equation}\label{4Potential1}
V(x)=
\begin{cases}
& 20 meV, \quad x\in R_2,\\
& V_g(x), \quad x\in R_3,\\
& 20 meV, \quad x\in R_4.
\end{cases}
\end{equation}
The potential mainly includes three parts, two barriers and a geometric potential region. In the rest two parts $R_1$ and $R_5$, the potential vanishes. Here $V_g(x)$ stands for the geometric potential~\eqref{GP} and $meV$ denotes milli electron volts.

To learn the action of the composite corrugation on the electronic transport, in what follows we will study the transmission probability of electron in the model Fig.\ref{Model} by the transfer matrix technique~ \cite{Ando1987Calculation}. In the models including a single-frequency corrugation~\cite{Encinosa2000Surface, Wang2016Transmission}, the boundaries constructed by two different materials, in which electrons have different effective masses, create resonant tunneling peaks and valleys. These results are still valid to the case of double-frequency corrugation. Although the single-frequency corrugation has been discussed widely~\cite{Shima2009Geometry, Shima2009Tuning, Wang2016Transmission}, the double-frequency corrugation (see Fig.~\ref{Model}) is still rarely discussed. Now we pay our attentions on the geometric effects of double-frequency corrugation on the transmission probability. We consider a model which consists of two barriers and a thin film with double-frequency corrugations.

\begin{widetext}

\begin{figure}[htbp]
\includegraphics[width=0.91\textwidth]{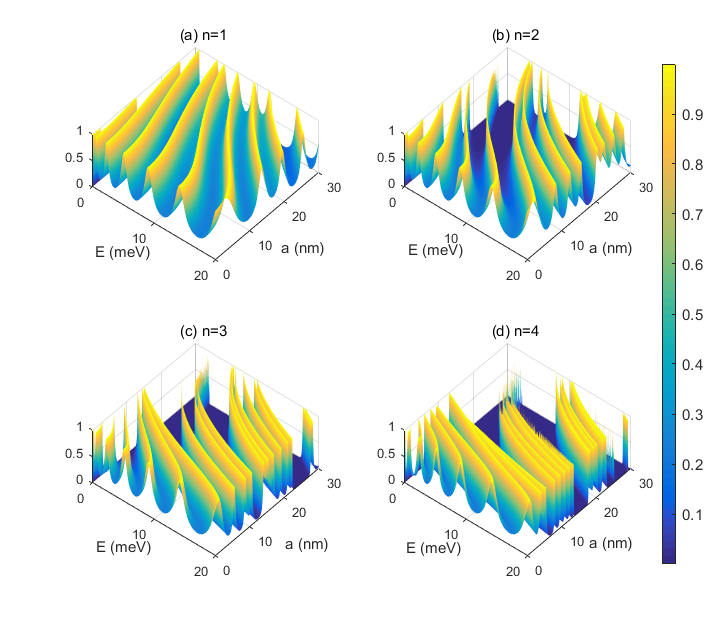}
\caption{\footnotesize (Color online) Transmission probability versus incident energy $E$ and the amplitude $a$ with $V(x)=20meV$ in $R_2$ and $R_4$, $V(x)=0meV$ otherwise, $m^*=0.067m_0$ in $R_3$ and $m^*=m_0$ otherwise, for (a) $n=1$, (b) $n=2$, (c) $n=3$ and (d) $n=4$.}\label{Trans3d}
\end{figure}

\end{widetext}

In the case of a surface reconstructed by double-frequency corrugations, the geometric potential $V_g$ can be exactly determined as Eq.~\eqref{GP}. And we can then calculate the transmission probability by using the transfer matrix method with $m^*=0.067m_0$ in $R_3$ and otherwise $m^*=m_0$. Subsequently, the transmission probability as a function of the incident energy $E$ and the amplitude $a$ of the double-frequency corrugation is described in Fig.~\ref{Trans3d} for (a) $n=1$, (b) $n=2$, (c) $n=3$ and (d) $n=4$, respectively. It is straightforward to see that there are resonant tunneling peaks and valleys of transmission probability, which are mainly created by the boundaries constructed by two different materials. When the multiple frequency $n=0$, the result is in full agreement with that in Ref.~\cite{Wang2016Transmission}. As $n=1$, the result is basically equivalent to that of single-frequency corrugation in the form. In striking contrast to the known results of the single-frequency corrugation, there is a fascinating result, the transmission gaps become wider and tunneling peaks become sharper for the cases of $n=2,3,4$. The manifest phenomena is entirely determined by the composite corrugation, because that the composite corrugation generates a geometric potential which consists of attractive wells with different depth. The reason is that electrons are scattered by those attractive wells, as a geometric scattering~\cite{Oflaz2018Scattering}. In other words, these interesting results are essentially from the geometry of composite corrugation introduced in $R_3$ of the present model. The multiple frequency $n$ is larger, the transmission gaps are rapidly becoming wider and the resonant tunneling bands are manifestly becoming sharper.

\begin{figure}[htbp]
\includegraphics[width=0.49\textwidth]{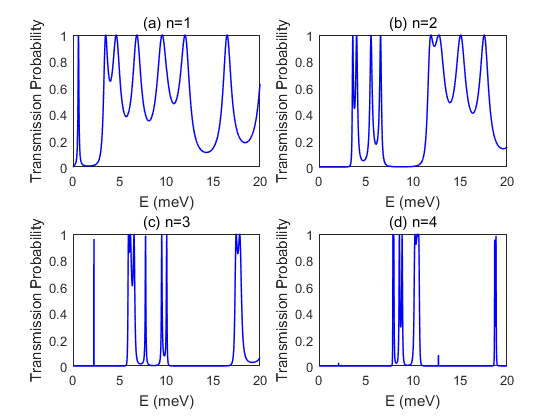}
\caption{\footnotesize (Color online) Transmission probability versus incident energy $E$ with $a=30nm$ and $m^*=0.067m_0$ in $R_3$, $m^*=m_0$ otherwise, for (a) $n=1$, (b) $n=2$, (c) $n=3$ and (d) $n=4$.}\label{Trans2d}
\end{figure}

Mathmatically, the transport distance is a function of the multiple frequency $n$ and the amplitude $a$ of the composite corrugation which can be expressed by
\begin{equation}\label{XiLength}\nonumber
\xi=\int_{R_3}\sqrt{1+a^2\gamma^2(\cos n\gamma x\cos\gamma x-n\sin n\gamma\sin\gamma x)^2}dx.
\end{equation}
According to the integral, the distance between adjacent wells obviously grows with increasing the amplitude $a$, while the distance is affected by growing the multiple frequency $n$ not obvious. Strikingly, the transmission gaps are effectively widened by growing $n$ with a fixed $a$, the tunneling domains are manifestly narrowed and sharpened (see Fig.~\ref{Trans2d}). The reason is that as the value of $n$ increases, the attractive wells rapidly becomes deeper and more. It is striking that the transmission gaps and tunneling domains are mainly contributed by the double-frequency corrugation. As a potential application, we can use composite corrugation to design an electronic switch, the transmission gaps mean that electron is reflected, but the tunneling domains do that electron can pass.

\section{Conclusions}\label{4}
In the present paper, we have considered a 2D thin film including a particular part that is a surface reconstructed by double-frequency corrugations. The particular corrugation generates a geometric potential, which consists of many attractive wells with different depth. Due to the presence of two frequencies, the attractive wells have not identical depth. The diversity of the well depth depends entirely on the multiple frequency $n$. The depth of wells grows by increasing the amplitude $a$, and extremely grows by increasing the multiple frequency $n$. Approximately, these attractive wells can be roughly replaced by square wells. The square wells with different depth can be structured by introducing periodically magnetic fields with different strengthes. By using magnetic field determined by magnetic vector potential, the filter designed for electron with certain energy is named the vector-tunable filter~\cite{Anna2009Wave}. By designing the composite corrugation of film, the filter fabricated for electron with certain energy can be named as a geometry-tunable filter. In comparison to the single frequency corrugation, the double-frequency corrugation effectively provides more accesses to adjust the electronic filter. As a particular application, the composite corrugation can used to design curvature-tunable electronic switch when the multiple frequency $n$ takes $n=3, 4$ and so on.

\section*{Acknowledgments}
This work is jointly supported by the Natural Science Foundation of Shandong Province of China (Grant No. ZR2017MA010), the National Major state Basic Research and Development of China (Grant No. 2016YFE0129300), the National Nature Science Foundation of China (Grants No. 11690030, No. 11475085, No. 11535005, No. 61425018).

\bibliographystyle{apsrev4-1}
\bibliography{myreferences}

\end{document}